\documentclass{emulateapj}\usepackage{apjfonts}
\def\Sref#1{\S\,\ref{sec:#1}}
\def\Fref#1{Fig.~\ref{fig:#1}}
\def\Tref#1{Table~\ref{tab:#1}}

\newcommand{\psr}{PSR B1957+20}
\begin{document}

\title{Evidence for a Massive Neutron Star from a Radial-Velocity
  Study of the Companion to the Black Widow Pulsar PSR B1957+20}
\shorttitle{Radial Velocity Study of the Companion to PSR B1957+20}

\author{M. H. van  Kerkwijk and R. P. Breton}
\affil{Department of Astronomy and Astrophysics, University
  of Toronto, 50 St.\ George Street, Toronto, ON M5S 3H4, Canada;
  mhvk,breton@astro.utoronto.ca}
\author{S. R. Kulkarni}
\affil{Palomar Observatory, California Institute of Technology
       105-24, Pasadena, CA 91125, USA; srk@astro.caltech.edu}

\slugcomment{Accepted for publication in the Astrophysical Journal 13 December 2010}
\begin{abstract}
  The most massive neutron stars constrain the behavior of ultra-dense
  matter, with larger masses possible only for increasingly stiff
  equations of state.  Here, we present evidence that the black widow
  pulsar, \psr, has a high mass.  We took spectra of its strongly
  irradiated companion and found an observed radial-velocity
  amplitude of $K_{\rm obs}=324\pm3{\rm\,km\,s^{-1}}$.  Correcting
  this for the fact that, due to the irradiation, the center of light
  lies inward relative to the center of mass, we infer a true
  radial-velocity amplitude of $K_2=353\pm4{\rm\,km\,s^{-1}}$ and a
  mass ratio $q=M_{\rm PSR}/M_2=69.2\pm0.8$.  Combined with the
  inclination $i=65\pm2\,$deg inferred from models of the lightcurve,
  our best-fit pulsar mass is $M_{\rm PSR}=2.40\pm0.12\,M_\odot$.  We
  discuss possible systematic uncertainties, in particular in the
  lightcurve modeling.  Taking an upper limit of $i<85\,$deg based on
  the absence of radio eclipses at high frequency, combined with a
  conservative lower-limit to the motion of the center of mass,
  $K_2>343{\rm\,km\,s^{-1}}$ ($q>67.3$), we infer a lower limit to the
  pulsar mass of $M_{\rm PSR}>1.66\,M_\odot$.
\end{abstract}

\section{Introduction}

One of the outstanding problems in physics is the behavior of matter
at extreme densities.  This behavior is modelled using
quantum-chromodynamics calculations, but these cannot yet determine
reliably the densities at which, e.g., meson condensation and the
hadron to quark-gluon phase transition occur.  At densities slightly
above nuclear and high temperature, models can be tested with
heavy-nuclei collision experiments.  For higher densities and low
temperatures, only comparison with neutron-star parameters is possible
(for a review, e.g., \citealt{lattp07}).  

The different models lead to different equations of state, which
predict different mass-radius relations for neutron stars.
Unfortunately, most attempts at observational tests have been
frustrated by susceptibility to systematic errors and modeling
uncertainties.  The most robust tests have involved measurements of
extrema.  For instance, the fastest measured spin period, 1.4\,ms (Ter
5ad; \citealt{hess+06}), excludes the stiffest equations of state, for
which neutron stars would be too large to spin so fast.  

A problem with measurements of the extrema, is that whether they occur
in Nature depends not only on whether they are allowed physically, but
also whether they are expected astronomically.  From models of stellar
evolution, \citet{timmww96} find neutron-star mass distributions at
birth with two narrow peaks, at $1.3$ and $1.8\,M_\odot$, containing
remnants of stars with initial masses smaller and larger than
$\sim\!19\,M_\odot$, respectively.  In binaries, however, where much
of the stellar envelope is removed during the evolution, they expect
only to form neutron stars in the lower mass bin.  The above may
explain why until recently most accurate masses were all close to
$1.4\,M_\odot$: most were measured for binaries containing pulsars
with neutron star companions, where the preceding evolution predicts
relatively little mass has been accreted (for a review,
\citealt{stai04}).  The one exception was the X-ray binary Vela X-1,
for which a higher mass of $1.86\pm0.16\,M_\odot$ was inferred
\citep{barz+01,quai+03}.  Such a high mass would imply, e.g., that
meson condensation is not important.  However, the large uncertainty
prevented a definitive conclusion.

However, neutron stars can become more massive after birth by
accretion.  Accretion also leads to ``recycling'' of radio pulsars: it
increases their spin frequency and decreases their magnetic fields.
Most recycled pulsars are accompanied by low-mass, $\la\!0.2\,M_\odot$
companions.  For these systems to form in the age of the Universe, the
companion must originally have been a star of $\ga\!0.8\,M_\odot$.
Thus, the companion lost almost all of its mass, and some of it should
have landed on the neutron star, leading to a concomitant increase in
the neutron-star mass.  Initial tests, however, were suggestive but
not conclusive: for pulsars with low-mass white dwarf companions,
masses around 1.4--1.7\,$M_\odot$ were found
\citep{vkerbk96,jaco+05,bass+06,verb+08}, somewhat larger than
typical, but not yet very constraining.  Presumably, in these systems
much of the mass actually left the binary.

More recently, higher masses were found in different types of
binaries, starting with a number of pulsar binaries in globular
clusters \citep{rans+05,frei+08a,frei+08b}.  In these cases, however,
the masses rely on observations of periastron advance, which is
assumed to be due to general relativistic effects only (rather than
classical ones such as due to rotationally and tidally induced
quadrupoles), and statistical arguments that the inclinations are
unlikely to be very low.  Thus, it was still possible to doubt that
very massive neutron stars could exist.  

While we were writing and revising this work, however, such doubts
disappeared, with accurate mass determinations for \object{PSR
  J1614-2230} ($1.97\pm0.04\,M_\odot$, \citealt{demo+10}) and
\object{PSR J1903+0327} ($1.67\pm0.02\,M_\odot$), both relying on
measurements of Shapiro delay, which is not easily mimicked by other
processes.  These masses exclude many of the soft equations of state,
such as Kaon condensation as envisaged by \cite{browb94}.

Intriguingly, both of the above systems do not have low-mass
white-dwarf companions like most binary pulsars, but rather a more
massive, carbon-oxygen white dwarf (PSR~J1614-2230) and a solar-mass
main sequence star (PSR J1903+0327).  Thus, both systems also had
different evolutionary histories (puzzling for PSR J1903+0327; see
\citealt{frei+10}).  In this sense, our approach is similar, in that
we specifically target another group of binary pulsars with different
properties and evolutionary histories, and, therefore, perhaps
different masses, the so-called ``black widow pulsars.''

In black-widow systems, a millisecond pulsar is accompanied by a
low-mass, few $0.01\,M_\odot$ companion, which is bloated and strongly
irradiated by the pulsar, leading to outflows strong enough to eclipse
the pulsar signal for significant fractions of the orbit.  The
irradiation causes strong heating on the side of the companion facing
the pulsar, and, as a result, strong orbital brightness variations of
the optical counterparts \citep{kulkdf88,vpar+88,stapbb96}.  From
detailed modelling of the lightcurves, the inclinations can be
constrained \citep{callvpr95,stap+99,reyn+07}, which, when combined
with velocity information, can be used to derive masses.

Here, we present a radial-velocity study of \psr\ \citep{frucst88},
the proto-type black-widow system, which has the brightest and
best-studied counterpart.  In \Sref{obs}, we describe our observations
and data reduction, and in \Sref{spectral} we constrain the
properties of the companion from the spectra, determine radial
velocities, and fit an observed radial-velocity amplitude.  In
\Sref{radinc}, we discuss the available constraints on the radius and
inclination, and in \Sref{corrections} the corrections we need to make
because our velocity amplitude is that of the center of light, which,
because of the irradiation, is shifted towards the pulsar relative to
the center of mass.  We present our final constraints on the masses in
\Sref{masses} and discuss how these may be made more secure in the
future.

\begin{deluxetable}{l@{\hskip1mm}l@{\hskip2mm}rlllll}
\tablewidth{0pt}
\tablecaption{Log of Observations and Velocity
  Measurements\label{tab:log}}
\tablehead{
&&\colhead{$t_{\rm int}$}&&&&&\colhead{$v_{\rm b}$}\\
\colhead{Date}&\colhead{UT}&\colhead{(s)}&\colhead{MJD$_{\rm 
mid,bar}$}&
\colhead{$\phi$}&\colhead{$f_{\rm b}$}&\colhead{$f_{\rm r}$}&
\colhead{$({\rm km\,s^{-1}})$}\\
\colhead{(1)}&\colhead{(2)}&\colhead{(3)}&\colhead{(4)}&
\colhead{(5)}&\colhead{(6)}&\colhead{(7)}&\colhead{(8)}
}
\startdata
2007 Jun 15& 11:41&  930 & 54266.48677& 0.55& 0.65& 0.74&   \phs$419\pm\phn6$\\ 
           & 11:58&  930 & 54266.49844& 0.58& 0.81& 0.90&   \phs$405\pm\phn5$\\ 
           & 14:40&  930 & 54266.61091& 0.87& 1.13& 1.08&\phn   $-97\pm\phn6$\\ 
           & 14:54&  640 & 54266.62052& 0.90& 0.94& 1.07&      $-139\pm   22$\\ 
2008 Aug 4 & 06:26& 1500 & 54682.26797& 0.08& 0.18& 0.29&      $-179\pm   13$\\ 
           & 10:12& 1800 & 54682.42481& 0.49& 0.41& 0.50&   \phs$429\pm\phn6$\\ 
           & 10:42& 1800 & 54682.44608& 0.54& 0.64& 0.69&   \phs$422\pm\phn5$\\ 
           & 11:16& 1800 & 54682.46951& 0.60& 0.90& 0.90&   \phs$385\pm\phn5$\\ 
           & 11:47& 1800 & 54682.49124& 0.66& 1.09& 1.05&   \phs$298\pm\phn5$\\ 
           & 12:12& 1000 & 54682.50830& 0.71& 1.23& 1.18&   \phs$212\pm\phn6$\\ 
           & 13:06&  900 & 54682.54576& 0.80& 1.31& 1.18&\phn\phs$22\pm\phn7$\\ 
           & 13:24& 1200 & 54682.55845& 0.84& 1.18& 1.11&\phn   $-48\pm\phn5$\\ 
           & 13:50& 1800 & 54682.57640& 0.88& 1.00& 1.02&      $-125\pm\phn6$\\ 
2008 Aug 5 & 08:37& 1800 & 54683.35900& 0.93& 0.75& 0.87&      $-184\pm\phn8$\\ 
           & 09:08& 1800 & 54683.38036& 0.99& 0.57& 0.63&      $-213\pm\phn9$

\enddata
\tablecomments{
  Col.\ (1): Date of the observation. 
  Col.\ (2): Start time.  
  Col.\ (3): Integration time.  
  Col.\ (4): Mid-exposure, barycentric Modified Julian Date.
  Col.\ (5): Phase using epoch of ascending node $T_0={\rm
    MJD~}48196.0635242$ and orbital period $P=33001.91484\,$
  (\citealt*{arzoft94}; we ignored orbital-period derivatives, see
  text). 
  Col.\ (6): Flux ratio relative to the contaminator in the
  5100--5300\,\AA\ range.
  Col.\ (7): Flux ratio in the 8450--8650\,\AA\ range.
  Col.\ (8): Radial velocity relative to the contaminator inferred
  from the blue spectra (we estimate a barycentric velocity of the
  contaminator of $-25.6\pm1.3{\rm\,km\,s^{-1}}$).}
\end{deluxetable}

\section{Observations \& Reduction}
\label{sec:obs}
We obtained pilot observations at the Keck telescope of the companion
of \psr\ on the night of 15 June 2007, and a larger set of spectra on
the nights of 4 and 5 August 2008 (see \Tref{log}).  For all
observations, the seeing was good, ranging from $0.6$ to
$\sim\!1\arcsec$, but the sky was not photometric.  For relative flux
calibration, we obtained spectra of a number of spectrophotometric
standards from \citet{bohlcf95} (see below).  We obtained exposures of
internal flat fields and Hg/Kr/Ar and Ne/Ar arc spectra interspersed
with the observations.

The spectra were obtained using the two-armed Low Resolution Imaging
Spectrometer (LRIS, \citealt{oke+95,mcca+98}).  We employed the
atmospheric dispersion corrector, used a $0\farcs7$ slit, set to
position angle $35\arcdeg$ to cover both the companion and a nearby
star (at $\sim\!1\farcs3$, hereafter the contaminator\footnote{The
  pulsar's proper motion has increased the separation over that quoted
  in earlier publications.}), and split the light with a dichroic at
6800\,\AA.  In the blue arm, we used the $600{\rm\,line\,mm^{-1}}$
grism, blazed at 4000\,\AA, which covers 3100--5600\AA\ at a
resolution $\Delta\lambda\simeq3.2\,$\AA\ or $\Delta
v\simeq220{\rm\,km\,s^{-1}}$ (for the $0\farcs7$ slit).  The detector
is a mosaic of two Marconi CCDs, each with $4096\times2048$ pixels of
$15\,\mu$m on the side ($0\farcs135$ on the sky), which we binned by
two in the dispersion direction (the only direction it can be binned).
On the red side, we used the $1200{\rm\,line\,mm^{-1}}$ grating,
blazed at 7500\,\AA, set to cover 7600--8900\,\AA\ at
$\Delta\lambda=2.1\,$\AA\ or $\Delta v\simeq75{\rm\,km\,s^{-1}}$.
Here, the detector was a Tektronix CCD with $2048\times2048$ pixels of
$24\,\mu$m on the side ($0\farcs215$ on the sky), which we read out
unbinned.

For the reduction, we used the Munich Image Data Analysis System
ESO-MIDAS, and routines running in the MIDAS environment.  For all
images, we subtracted bias as determined from the overscan regions.
For the blue images, we subsequently corrected for small-scale
variations in efficiency by dividing by a spatially averaged flat
field, normalized using a third-degree polynomial, and with the
bluest, poorly exposed part shortward of 4000\,\AA\ replaced by unity.
For the red images, we simply divided by the flat field, normalized
using a bi-linear fit.  

The spectra and their uncertainties were extracted by fitting, at each
dispersion position, the sum of three stellar profiles (for the
companion, contaminator, and another star at $\sim\!3\farcs2$ on the
other side of the target) and a constant sky.  For the profiles, we
used Moffat functions of the form $P=A/(1+(x/w)^2)^\delta$, with power
$\delta=5$ for the blue images and $\delta=6$ for the red ones.  At
each dispersion position, only the three amplitudes $A$ and the sky
were fitted; the other parameters were determined globally (with the
central position and width $w$ allowed to vary quadratically with
dispersion position, and the relative positions fitted as constants).
The fits to the images were generally good for the pulsar fields, with
reduced $\chi^2$ near unity, and somewhat poorer for the higher
signal-to-noise images of the spectrophotometric standards (where
slight mismatches between the Moffat function and the true point
spread function are more apparent).

Wavelength calibration was done using arc spectra.  For the blue arm,
we used well-exposed images of the arc lamps taken at the start of the
night to define an overall solution, which required a fourth-order
polynomial to give an adequate dispersion solution, with
root-mean-square residuals of 0.14\,\AA\ for 24 lines.  Next, we found
offsets relative to this solution from less well-exposed arc frames
taken interspersed between the observations.  For the red arm, the
individual arc frames were well exposed, and a third-degree polynomial
sufficed to give solutions with residuals of $\sim\!0.04\,$\AA\ (for
typically $31$ lines).

For flux calibration, we first corrected all spectra approximately for
atmospheric extinction using a curve made by combining the CFHT values
\citep*{belabd88} shortward of 5200\,\AA\ with the better sampled La
Silla values longward of 5200\,\AA\ (ESO users manual 1993; see also
\citealt{tug77}).  Next, for the blue spectra, we calculated response
curves by comparing our observed spectra for Feige 110 with the
calibrated STIS spectra given by \citet{bohlcf95}: we slightly
smoothed our spectra to match the STIS resolution, divided the two,
and smoothly interpolated the ratio.  For the 2007 red spectra, we
proceeded similarly, except that we compared our observation of BD +28
4211 with a third-degree polynomial fit to the STIS spectrum of
\citet{bohlcf95}, since the STIS spectrum was relatively noisy and the
intrinsic spectrum should be smooth.  Furthermore, we took care to fit
the ratio spectrum avoiding telluric absorption, so that we could use
the deviations -- scaled with airmass -- to correct to first order for
telluric absorption in our target spectra.  For the 2008 red spectra,
our procedure was similar, except that we compared our observed
spectrum of Hz 43A with the calibrated model spectrum (which is again
smooth over the relevant wavelength range; \citealt{bohlcf95}).  Since
the conditions were not photometric, the above gives good relative
fluxes over the observed wavelength range, but only approximate
absolute fluxes.  To place the companion spectra on the same flux
scale, we scaled all sets of spectra to the same average contaminator
flux in the 5100-5300\,\AA\ and 8450--8650\,\AA\ ranges (for blue and
red, respectively).  After this scaling, integrating over B filter
band passes, we find that our inferred companion magnitudes are
roughly consistent with the ones observed by \citet{reyn+07}.

\section{Spectra and Velocities}
\label{sec:spectral}

In \Fref{spectra}, we show averaged blue and red spectra for phases
within 0.15 from superior conjunction of the companion, when the
irradiated side is seen face on, and for phases between 0.15 and 0.3
from superior conjunction, when the view is more sideways.

Below, we first determine the spectral type for the two phases.  While
this will be only an ``average'' spectral type over the surface, it
gives a way to determine the influence of irradiation.  We will find
that the spectrum appears rather normal, suggesting that the
irradiated energy is dissipated well below the photosphere (in
contrast to the case of, e.g., the pre-cataclysmic variable NN~Ser,
where the companion is irradiated with ultraviolet light from a hot
white dwarf; see \citealt{pars+10}).  

Given the normal appearance of the spectra, we use them to help
determine reddening and distance, and to constrain the surface gravity
and radius of the companion.  In \Sref{corrections}, we will find that
these provide additional evidence that the companion is close to
filling its Roche lobe.  Next, we describe how we determined the
velocities from the individual spectra, and use these to fit an orbit.
We discuss the required corrections to this orbit in
\Sref{corrections}.

\begin{figure*}
\centering
\includegraphics[angle=-90,width=0.9\hsize]{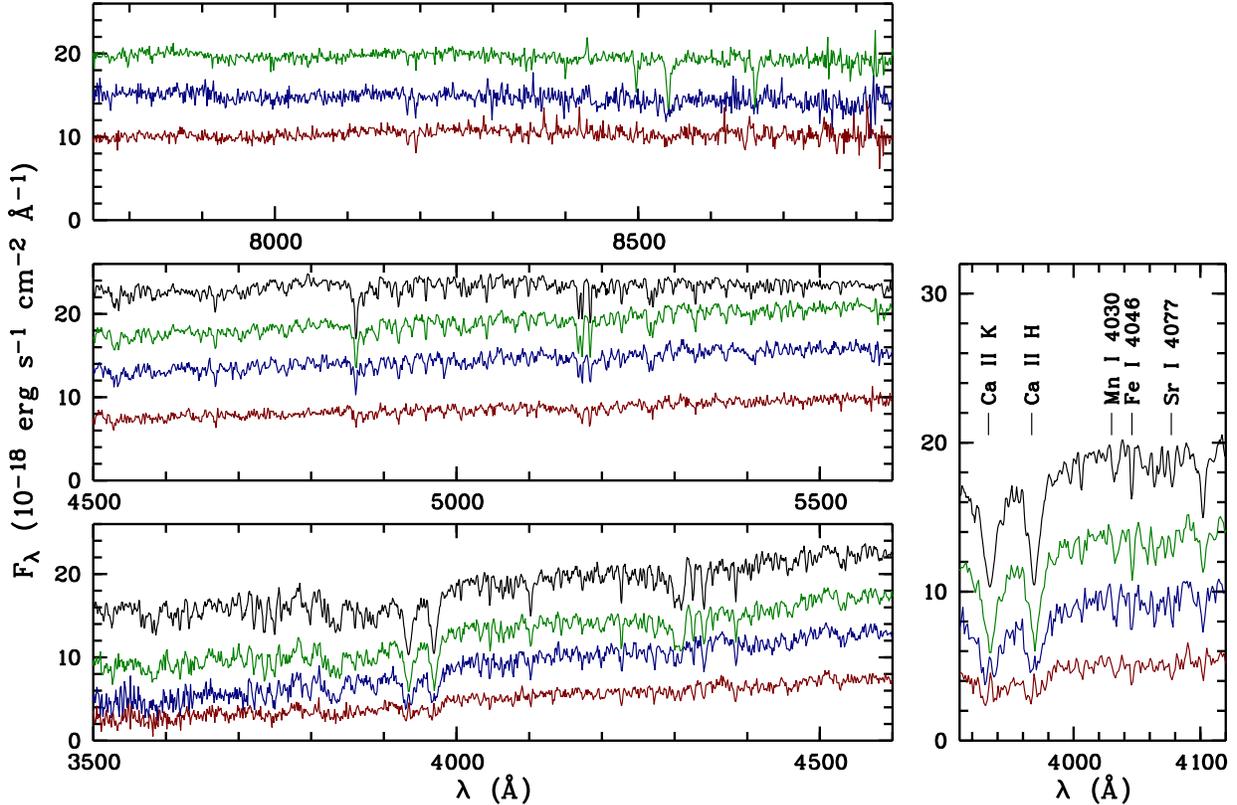}
\caption{Spectra of the companion to \psr.  We show averages both
  around superior conjunction (blue trace), when the illuminated side
  is most in view, and nearer quadrature (red trace).  Also shown are
  the spectrum of the nearby ``contaminator'' (green trace, offset by
  5 units), as well as a UVES spectrum of the G2\,IV star $\beta$~Hyi
  \citep{bagn+03} (black trace; convolved to the same resolution,
  scaled to match the brightness of the conjunction spectrum, and
  offset by 10 units; the red part of that spectrum is not shown, as
  it has residual ripples and does not cover the \ion{Ca}{2} triplet).
  The right-hand enlargement shows the \ion{Ca}{2} H and K lines,
  which have cores in emission for the companion, and lines of
  \ion{Sr}{2} at 4077\,\AA, \ion{Fe}{1} at 4046\,\AA, and \ion{Mn}{1}
  at 4030\,\AA, which we use to estimate the luminosity class and
  surface gravity.}
\label{fig:spectra}
\end{figure*}

\subsection{Spectral Type, Reddening, and Distance}
\label{sec:spectral_type}

We compared our blue spectra with classification spectra shown in the
on-line atlas by
R. O. Gray.\footnote{\url{http://nedwww.ipac.caltech.edu/level5/Gray/Gray\_contents.html}}  We find that the contaminator has spectral
type G1--2\,V, while that of the companion is slightly earlier at its
brightest phase (F9--G0, as seen from, e.g., the weaker G band near
4300\,\AA), and similar when seen from the side (about G1).  Its
luminosity class is slightly higher, intermediate between IV and III,
as can be seen from the stronger \ion{Sr}{2} $\lambda4077$ line, and
also supported by the lower strength of the \ion{Fe}{1} $\lambda4046$
line relative to the \ion{Mn}{1} $\lambda4030$ line.  Consistently, in
\Fref{spectra}, one sees that $\beta$~Hyi, which has spectral type
G2\,IV, is intermediate in luminosity class between the contaminator
and the companion.  For $\beta$~Hyi, the mean density and radius have
been measured using asteroseismology and interferometry, leading to a
surface gravity $\log{}g=3.952\pm0.005$ \citep{nort+07}.  For the
companion of \psr, we thus infer $\log{}g\lesssim4$.

The blue spectra of the companion show emission cores in the
\ion{Ca}{2} H and K lines, suggestive of an active chromosphere.
Furthermore, H$\beta$ seems rather weak, possibly due to being filled
in by poorly subtracted emission from the Balmer-dominated bow shock
nebula (\citealt{kulkh88}; the contaminator might also be affected by
this).  Finally, the \ion{Mg}{1}b triplet is a bit weaker than
expected.

In the red spectra, we find that the spectrum of the contaminator is
as expected for a G1--2\,V star, but that the spectra of the companion
cannot be classified as easily.  In particular, the \ion{Ca}{2} IR
triplet nearly absent, perhaps being filled in by chromospheric
emission.  Furthermore, the \ion{Na}{1} $\lambda\lambda$8183,8195
lines are much stronger than seen in the contaminator: combined
equivalenth widths of $\sim\!0.9$ and $1.5\,$\AA\ for the face-on and
more sideways view of the companion, respectively, and
$\sim\!0.7\,$\AA\ for the contaminator.  Probably, this reflects that
in the red, a large contribution to the light arises from cooler
regions of the companion, which will have much stronger \ion{Na}{1}
absorption (e.g., \citealt{zhou91} finds equivalent widths of
$\sim\!0.7\,$\AA\ and $\sim\!1.5\,$\AA\ for early G and early M
dwarfs, respectively).  The strength of the \ion{Na}{1} line also
shows that the surface gravity of the companion is similar to that of
a main-sequence or sub-giant star; for giants, the \ion{Na}{1}
equivalent width does not exceed 1\,\AA\ \citep{zhou91} for any
temperature.

We can use the spectral information to constrain the reddening to the
source.  Starting with the contaminator, from archival {\em Hubble
  Space Telescope} observations with WFPC2, which we analyzed with
HSTphot \citep{dolp00}, we measure $m_{\rm F675W}=19.968\pm0.010$ and
$m_{\rm F814W}=19.458\pm0.015$.  Using the transformations of
\cite{holz+95}, this corresponds to $R=20.06\pm0.03$ and
$I=19.40\pm0.02$.  Since a G1--2\,V star has $(R-I)_0=0.34\pm0.02$ and
$M_R\simeq4.3$ \citep{cox00}, we infer an extinction
$A_V=1.42\pm0.18\,$mag and distance of $\sim\!8\,$kpc (where we use
the extinction curve of \citealt*{schlfd98}).


The reddening to the companion to \psr\ should be similar to that to
the contaminator, since its observed colors at maximum are about as
much bluer (from our blue spectra, $\Delta(B-V)=-0.07\pm0.02$) as the
intrinsic color difference expected from the spectral type
($\Delta(B-V)_0=-0.05\pm0.02$).
The reddening can be compared with the run of reddening with distance
measured using red-clump stars.  Using the technique of
\citet{duravk06}, we find that the reddening is $A_V\simeq0.5$ at
$d\simeq1.5$\,kpc, increases first slowly and then more sharply to
$A_V\simeq1.2$ at $d\simeq2\,$kpc, and remains constant thereafter.
Given that the contaminator is certainly well beyond 2\,kpc, and that
the reddening of \psr\ and the contaminator are similar, we conclude
that \psr\ is at a distance $d\gtrsim2\,$kpc.  This is consistent with
the distance in the range of 1.5 to 2.5\,kpc inferred from the pulsar
dispersion measure of $29{\rm\,cm^{-3}\,pc}$ using models of the
Galactic electron distribution \citep{taylc93,cordl02}.  From the
observations of \cite{callvpr95}, the companion at maximum brightness
is about 0.4\,mag brighter than the contaminator in the R band, which,
using the R-band magnitude above,\footnote{From Fig.~1 of
  \cite{callvpr95}, we read off $R=19.89\pm0.03$ for the contaminator,
  with a quoted systematic uncertainty of 0.05\,mag.  This is somewhat
  brighter than what we measured from the {\em HST} data.  We use the
  fainter magnitude to obtain a conservative limit on the radius.}
implies $R=19.7$.  This corresponds to an absolute magnitude
$M_R\simeq7.2-5\log d_2$ and an effective radius
$R\simeq0.25d_2\,R_\odot$ (where $d_2\equiv d/{\rm2\,kpc}$).




\begin{figure}
\centering
\includegraphics[width=0.9\hsize]{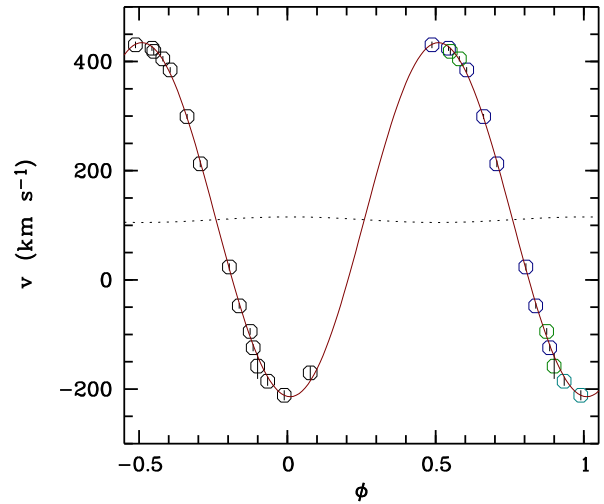}
  \caption{Radial-velocity measurements of the companion of \psr\ as a
    function of orbital phase.  The colored circles mark velocity
    measurements from different nights, with errors as indicated.
    They are repeated in black for clarity.  The drawn curve is the
    best-fit circular orbit, and the dotted one the pulsar orbit
    inferred from radio timing.}
\label{fig:rv}
\end{figure}

\subsection{Radial Velocities}
\label{sec:RV}

We determined velocities by fitting the flux-calibrated blue spectra
of both the pulsar companion and the contaminator with a template
based on the high-resolution spectrum of the G2\,IV star $\beta$~Hyi
(HD~2151) from the UVESPOP library \citep{bagn+03}.  For our fitting,
we convolved the UVES spectrum with a truncated Gaussian to match the
resolution of the observed spectra corresponding to the
$\sim\!0\farcs7$ seeing and $0\farcs7$ slit.  We ignored rotational
and orbital broadening, which are well below our resolution for the
blue spectra and just comparable to that of the red spectra (for
companion radius $R_{\rm c}\simeq0.25\,R_\odot$, one finds $v\sin
i=2\pi R_{\rm c}\sin i/P\simeq30{\rm\,km\,s^{-1}}\sin i$; for
integration time $t_{\rm int}=1800\,$s, the maximum smearing is
$K_{\rm obs}\sin(2\pi t_{\rm int}/P)\simeq 80{\rm\,km\,s^{-1}}$, where
$K_{\rm obs}=324{\rm\,km\,s^{-1}}$ is the radial-velocity amplitude
[see below]).

The template was fitted for a grid of velocities between $-600$ and
$+600\,{\rm\,km\,s^{-1}}$ with a step size of $5{\rm\,km\,s^{-1}}$, at
each velocity fitting for the normalization and possible variation
with wavelength using a quadratic function.  Typical reduced $\chi^2$
values for the best fits were $\chi^2_{\rm red}\simeq1.2$.  The best
fit velocity was determined using a quadratic fit to the $\chi^2$
values within $40{\rm\,km\,s^{-1}}$ of the minimum.  Looking at the
results for the contaminator, it is clear that systematic variations
are present, with root-mean-square of $13{\rm\,km\,s^{-1}}$, likely
due to small shifts in placement in the slit and/or uncorrected
atmospheric dispersion.  As our velocities, we thus take the
difference between the velocities inferred for the companion and the
contaminator; these velocities and their corresponding uncertainties
are listed in \Tref{log}.

To see how sensitive our results are to our choices, we tried fitting
only part of the blue spectrum, and using different UVESPOP stars or
model atmospheres as templates.  We found that the velocities were
consistent to within $\sim\!1\sigma$, which we will use as an estimate
of the associated systematic uncertainty below.  We also tried fitting
the red spectra, but found that for most, the absence of the expected
lines combined with the poor signal-to-noise did not allow us to
obtain a reliable velocity.

We fitted the velocities with a circular orbit using epoch of
ascending node $T_0={\rm MJD~}48196.0635242$ and orbital period
$P=33001.91484\,$s \citep*{arzoft94}, but found we could obtain a good
fit only if we left the phase free (\Fref{rv}).  Compared to the
prediction, ascending node occurs $\sim\!350\,$s later (phase offset
$0.011\pm0.002$).  However, the orbital period is known to vary
quasi-periodically (\citealt{arzoft94}; \citealt*{niceat00}); indeed,
from the first and second orbital period derivatives measured by
\citet{arzoft94}, the expected phase offset is $0.0406$.  Keeping the
phase free, our fit has $\chi^2_{\rm red}=0.92$ for 12 degrees of
freedom (15 measurements, 3 parameters).  The observed radial-velocity
amplitude is $K_{\rm obs}=324\pm3{\rm\,km\,s^{-1}}$ and the systemic
velocity, measured relative to the contaminator, is $\gamma_{\rm
  obs}=110\pm5{\rm\,km\,s^{-1}}$ (here, we multiplied the formal
errors with $\sqrt{2}$ to account for the systematic uncertainty in
the velocities related the choice of template and fitting region
discussed above).

The above systemic velocity is relative to the velocity of the
contaminator.  We tried to measure the latter in two different ways.
First, a simple average of the velocity measurements from the blue
spectra yields $-39\pm4{\rm\,km\,s^{-1}}$ (where we corrected for the
velocity of $22.7\pm0.9{\rm\,km\,s^{-1}}$ of $\beta$~Hyi,
\citealt{evan67}; we verified that we obtained the same result within
$\lesssim\!2{\rm\,km\,s^{-1}}$ using other UVESPOP stars and model
atmospheres).  Possible evidence that this is not reliable, however,
comes from the large scatter in the velocities from individual spectra
(see above).  A similar scatter is seen in the \ion{O}{1} 5577 sky
emission line.  The latter also shows an average offset, of
$6{\rm\,km\,s^{-1}}$, but it is not clear whether one can apply this,
since the line is very close to the red edge of the blue spectra,
where the wavelength solution may be less reliable.

As an alternative, therefore, we measured the velocities of the
contaminator using the red spectra, where, given the higher
resolution, any shifts due to offsets in the slit and/or atmospheric
dispersion should be smaller.  Here, we unfortunately could not use
the UVESPOP spectra, since these have a gap over the \ion{Ca}{2} IR
triplet, and thus we determined velocities relative to a $T_{\rm
  eff}=6000\,$K, $\log{}g=4.5$ model spectrum from \cite{zwitcm04}.
We find that the scatter in these velocities is substantially smaller
(root-mean-square of $6{\rm\,km\,s^{-1}}$), and that the (weighted)
average velocity is $-25.1\pm1.5{\rm\,km\,s^{-1}}$.  From the OH sky
line at 8344.602\,\AA, we infer that any offset due to wavelength
calibration errors are small; including these, we find an average
velocity of $-25.6\pm1.3{\rm\,km\,s^{-1}}$.

As a check on these numbers, we can compare the velocities with what
is expected for a star that is following simple Galactic rotation.
Assuming a flat rotation curve with $\Theta=220{\rm\,km\,s^{-1}}$ and
a distance to the Galactic center of 8.5\,kpc \citep{cox00}, as well
as a peculiar velocity of the Sun relative to the local standard of
rest of $(U,V,W)=(10.00,5.25,7.17){\rm\,km\,s^{-1}}$ \citep{dehnb98},
we find that for distances of 6, 8, and 10\,kpc for the contaminator,
the expected radial velocities are $+16.7$, $-1$, and
$-24{\rm\,km\,s^{-1}}$, consistent with our measurements.  (Arguably,
one should consider asymmetric drift in our estimates, in which case
the velocities would be more negative by $\sim\!15{\rm\,km\,s^{-1}}$.)

Overall, we believe the contaminator velocity from the red spectra is
more reliable.  For \psr, the implied systemic velocity is
$84\pm5{\rm\,km\,s^{-1}}$.




\section{Radius and Inclination}
\label{sec:radinc}

The companion is irradiated by the pulsar and presents a hot and a
cold side.  Hence, the center of light does not coincide with the
center of mass, but is shifted somewhat towards the pulsar.  As a
result, the spectroscopic observations underestimate the true
radial-velocity amplitude.  The correction depends on the stellar
radius, the temperature distribution, and the inclination.  Here, we
discuss the available constraints on these properties.  

The lightcurve provides strong constraints on the system parameters,
especially if simplifying assumptions can be made for the temperature
distribution.  For \psr, high-quality lightcurves were presented by
\cite{callvpr95} and \cite{reyn+07}.  These authors also fit their
lightcurves with models.

In these lightcurve synthesis models, it is assumed that the
companion's shape is that of an equipotential surface, and that its
temperature distribution is given by some background temperature
(modified suitably by gravity darkening) that is increased by
irradiation by an isotropic pulsar wind such that the outgoing flux
equals the sum of the background and irradiation fluxes (for a
detailed description, see \cite{orosh00}, who wrote the ELC code used
by \cite{reyn+07}).  It is assumed the irradiation does not affect the
temperature structure of the atmosphere (the ``deep heating''
approximation), such that each surface element can be taken to radiate
as predicted by a model atmosphere for a single star.  The main free
parameters are the background temperature and the irradiating flux
(which set the temperature distribution), the extent to which the
companion fills its Roche lobe (which determines its shape), and the
inclination of the orbit.

\cite{reyn+07} find that such models reproduce the lightcurves in
detail, and they infer that the companion nearly fills its Roche lobe,
up to a filling factor $R_{\rm nose}/R_{\rm L1}$ in the range
$0.81<R_{\rm nose}/R_{\rm L1}<0.87$ (where $R_{\rm nose}$ is the
radius of the star in the direction of the pulsar, and $R_{\rm L1}$
the distance to the inner Lagrangian point); this corresponds to a
volume-equivalent radius $R_2$ in the range $0.946<R_2/R_{\rm
  RL}<0.974$ (where $R_{\rm RL}$ is the volume-equivalent radius of
the Roche lobe).  They also infer an inclination $i$ in the range
$63^\circ<i<67^\circ$.

Formally, the above ranges are at the 3$\sigma$ level \citep{reyn+07}.
This, however, does not take into account uncertainties in the models
or the extent to which the underlying assumptions hold.  For instance,
the models may have temperature-dependent missing opacities, the
metallicity or hydrogen abundance may not be solar, the pulsar wind
that irradiates the companion may not be isotropic (though the
lightcurve is remarkably symmetric), or the deep heating approximation
may not be valid (although our blue spectra suggest it is not bad).
Given these issues, we will treat the ranges as 1$\sigma$
uncertainties below.

In addition to the above, an uncertainty that is more difficult to
constrain relates to the extent to which heat is redistributed.  Any
redistribution would lead to a smoother temperature distribution, and
thus it would require a larger inclination to obtain the same observed
modulation amplitude.  In principle, it should be possible to
constrain the temperature distribution directly by fitting multi-band
lightcurves and spectra simultaneously, similar to what has been done
for \object{NN Ser} by \citet{pars+10}, which is also strongly
irradiated (and for which no evidence for heat transport was found).
We hope to pursue this in the future.

Here, we will try to use the observations to set limits.  For our
purposes, the most important quantities are the companion radius, which
determines the correction to the radial-velocity amplitude (see
below), and the inclination, on which the final masses depend as
$1/\sin^3i$.

For the radius, our spectra yield independent clues.  First, the
surface gravity is limited to $\log g_2\lesssim4$.  Since the minimum
companion mass is $M_{2,\rm min}=0.022\,M_\odot$ (from the pulsar mass
function and our observed radial-velocity amplitude), one infers
$R_{2,g}=(g_2/GM_2)^{1/2}\gtrsim0.25\,R_\odot$.  Similarly, from the
distance limit inferred from the reddening, we found $R_{2,d}>R_{\rm
  eff}\gtrsim0.25\,R_\odot$.

These radii can be compared with the radius of the Roche lobe.  For a
companion much less massive than the pulsar, $R_{\rm
  RL}\simeq0.46a(M_2/[M_1+M_2])^{1/3}$ \citep{pacz71}.  The scaling
yields the well-known result that the mean density of the Roche lobe
is determined just by the period; numerically, $\overline\rho_{\rm
  RL}= 0.185{\rm\,g\,cm^{-3}\,}(P/{\rm1\,d})^{-2}$ \citep{eggl83}.
For \psr, $\overline\rho_{\rm RL}= 1.27{\rm\,g\,cm^{-3}}$, and hence
the size of the Roche lobe is $R_{\rm RL}=
(3M_2/4\pi\overline\rho_{\rm RL})^{1/3}=0.29\,R_\odot$ (where the
numerical value is for the minimum mass).  Thus, we conclude that
$R_{2,g}/R_{\rm RL}\gtrsim0.86$.  This is an overall lower limit,
since $R_{2,g}/R_{\rm RL}\propto M_2^{1/6}$.  Our limit is consistent
with what was inferred from the lightcurve fit (as well as from
theoretical considerations of the cause of orbital period variations;
\citealt{appls94}).  It implies a lower limit $R_{\rm nose}/R_{\rm
  L1}\gtrsim0.7$, which we will use below.


For the inclination, it is more difficult to set stringent limits.
The fact that at low radio frequencies, a symmetric eclipse is seen
that lasts $\sim\!8$\% of the orbit \citep{fruc+90,rybat91}, while no
eclipse is seen at high frequencies \citep{frucg92}, suggests that the
conditions along the line of sight do not change too strongly during
the eclipse.  This seems easier to understand if the line of sight
does not pass close to the companion, i.e., if the inclination is
intermediate, $i\simeq\arccos(R_E/a)\simeq75^\circ$ (for fractional
eclipse radius $R_E/a\simeq0.08\pi$), consistent with the inference
from the lightcurve.  Strictly, however, the eclipses only set a weak
upper limit: the absence of eclipses at high frequencies implies
$i<\arccos(R_2/a)\lesssim85^\circ$.

A lower limit to the inclination can be set from the large brightness
contrast between superior and inferior conjunction (a factor~100 in
the R band, \citealt{reyn+07}), which shows that at inferior
conjunction at most a small part of the irradiated hemisphere is
visible.  Conservatively, we estimate that this requires $i>50^\circ$.
A similar lower limit seems reasonable from the fact that the radio
eclipses last long; for lower inclinations, it is difficult to
envisage a physical eclipse region that does not extend all the way to
the pulsar.

Overall, we conclude that all observations are consistent with the
model inferences of \cite{reyn+07}, with $R_{\rm nose}/R_{\rm
  L1}=0.84\pm0.03$ and $i=65\pm2\,$deg.  Considering possible
systematic uncertainties, a secure constraint on the radius seems to
be $0.7<R_{\rm nose}/R_{\rm L1}<1$, while for the inclination the
constraint is $50<i<85\,$deg.

\section{Motion of the Center of Mass}
\label{sec:corrections}

The ratio between the observed radial-velocity amplitude of the center
of light and the actual one of the center of mass can be written as
$K_{\rm obs}/K_2=1-f_{\rm eff}R_{\rm nose}/a_2$, where the effective
normalized emission radius $f_{\rm eff}$ is constrained to be between
0 (uniform emission) and 1 (emission from the tip of the star facing
the pulsar only).  It should depend primarily on the surface
brightness and line strength distributions in the observed band, with
minor contributions arising from the exact shape of the star
(determined by the mass ratio, filling factor, and the degree of
co-rotation) and the orbital inclination.  Indeed, for the somewhat
similar situation of irradiation-induced Bowen emission lines on a
Roche-lobe filling companion, \cite{munocm05} found that the
``K-correction'' factor depends mainly on mass ratio (which determines
the radius) and is nearly independent of inclination.

We investigated this hypothesis by adapting a lightcurve synthesis
code to deal with spectra (the code was used by \cite{stap+99} to
model pulsar irradiation, and is similar to that of \cite{orosh00}).
The code produces not just fluxes, but also synthetic spectra, by
summing Doppler-shifted spectra over all surface elements.  While we
do not yet have a suitable set of atmosphere models in hand to model
our observations reliably, we have used the code to estimate the
effect on the radial-velocity amplitude, by generating mock spectra
for different sets of binary parameters.  We generated spectra at the
same orbital phases as our observations, determined radial velocities
using the same method as done for the real observations (\Sref{RV}),
and fitted circular orbits using the same weights.\footnote{For a
  distorted, irradiated model, the predicted radial-velocity curve is
  not necessarily circular (see, e.g., the curve for NN~Ser of
  \citealt{pars+10}).  But any resulting systematic effects are
  corrected for by fitting circular orbits to observed and
  model velocities at the same phases and with the same weighting.}

At a fixed mass ratio, we found that the orbital inclination had
negligible effect on $f_{\rm eff}$, $\lesssim\!1$\% over the range
$50<i<90\,$deg.  The effects of the filling factor, for the range
$0.7<R_{\rm nose}/R_{\rm L1}<0.95$, were a bit larger, though still
small, at $\sim\!7\%$.  The influence of the precise values of
strength of the irradiation is $\sim\!6$\%, for the range of
irradiation strengths that give front-side temperatures consistent
with our spectral type (6000--6500\,K); the backside temperature does
not matter much, since it is constrained to be $\sim\!2900\,$K
\citep{reyn+07}, much too low to contribute B-band flux.  As expected,
with our choice of scaling with $R_{\rm nose}$, the effect of the mass
ratio is very small, $\lesssim\!1$\% for the range of values that are
able to yield the observed $K_{\rm obs}$.  Overall, we infer
$f_{\rm eff}\simeq0.60\pm0.04$.

For comparison, looking from the side at a spherical star that is dark
on one side and emits isotropically at a uniform temperature on the
other, simple integration yields $f_{\rm eff}=4/3\pi=0.42$.  Assuming
instead a temperature distribution $T=T_0\cos^{1/4}\theta$, as
expected for irradiation by a parallel beam, we find $f_{\rm
  eff}=0.62$ for $T_0=6400\,$K (assuming black-body emission and a
linear limb darkening law with $u=0.6$).  This is very similar to what
we find from our model, showing that the increase in brightness
dominates over effects such as an increase in line strength with
decreasing temperature.

Before using the above to estimate $K_2$, it is useful first to
consider the limits.  For uniform emission, $K_2=K_{\rm obs} =
324\pm3{\rm\,km\,s^{-1}}$, which yields a lower limit to the mass
ratio for the system $q_{\rm min} = M_1/M_2 = K_{\rm obs}/K_1 =
63.6\pm0.6$ (where the pulsar's radial-velocity amplitude $K_1 = 2\pi
a_1\sin i/P_{\rm orb}=5.09272\pm0.00004{\rm\,km\,s^{-1}}$ is known
from radio timing; \citealt{arzoft94}).  At the other extreme, for
light emitted by the nose of a Roche-lobe filling star, i.e., $f_{\rm
  eff}R_{\rm nose}=R_{\rm L1}$, one finds $K_{\rm obs}=0.843K_2$,
i.e., $K_2=384\pm4{\rm\,km\,s^{-1}}$ and $q_{\rm max}=75.4\pm0.7$.  We
will take these values as limits in \Sref{masses} (accounting for the
fact that neither can be close to realistic by ignoring the
uncertainty on $K_{\rm obs}$).

For our more general case, combining our estimate of $f_{\rm
  eff}=0.60\pm0.04$ above with $R_{\rm nose}/R_{\rm L1}=0.84\pm0.03$,
and solving for $K_2$ (taking into account that $R_{\rm L1}/a$ depends
on $q$), one finds $K_2=353\pm4{\rm\,km\,s^{-1}}$.  For the
conservative range in companion radius, $0.7<R_{\rm nose}/R_{\rm
  L1}<1$, we find $348\pm4<K_2<358\pm4{\rm\,km\,s^{-1}}$, i.e., it
corresponds to a $5{\rm\,km\,s^{-1}}$ uncertainty in $K_2$.  Adding
this in quadrature to the 2$\sigma$ uncertainty of $8{\rm\,km\,s^{-1}}$
arising from the uncertainties in $K_{\rm obs}$ and $f_{\rm eff}$, we
infer a conservative range in radial-velocity amplitude
$343<K_2<363{\rm\,km\,s^{-1}}$.  From the above, we conclude that the
mass ratio is $q=69.2\pm0.8$, and that a conservative range is
$67.3<q<71.3$.

\begin{figure}
\centering
\includegraphics[width=0.9\hsize]{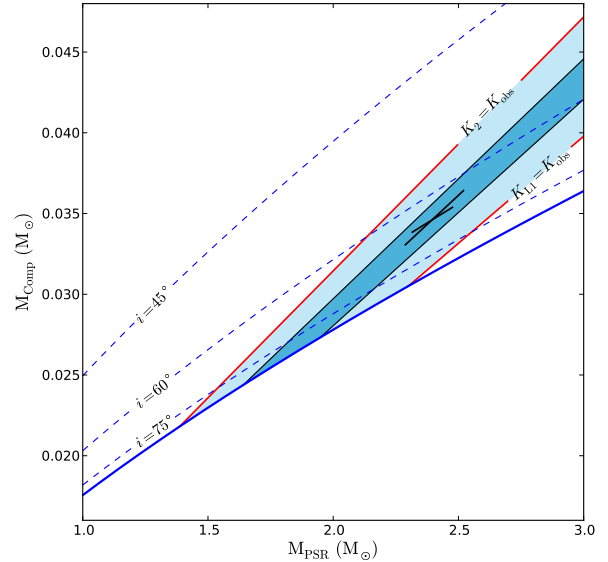}
\caption{Mass-mass diagram for \psr\ and its companion.  The cross
  indicates our best-fit solution with 1$\sigma$ uncertainties, and
  the surrounding blue parallelogram the convervative region including
  our best estimate of possible systematic uncertainties (see text).
  The physically allowed region (light blue) is limited by the
  constraints $\sin i\leq1$ (thick, blue line), and $K_{\rm L1}\leq
  K_{\rm obs}\leq K_2$ (thick, red lines).  Here, the second
  constraint arises because the observed radial-velocity amplitude
  $K_2$ is measured using light emitted from the side facing the
  pulsar, and the center of light cannot be further away from the
  pulsar than the center of mass, or closer to the pulsar than the
  first Lagrangian point~L1 (with velocity amplitudes $K_2$ and $K_{\rm
    L1}$, respectively).  For reference, we show contours of constant
  inclination~$i$ (dotted), calculated using the pulsar mass function.
}
\label{fig:masses}
\end{figure}

\section{Inferred Masses and Conclusions}
\label{sec:masses}

In \Fref{masses}, we show our constraints on the masses.  One sees
that most likely, \psr\ is massive, with $M_{\rm PSR}=2.40\,M_\odot$.
Taking the inferences from \citet{reyn+07} on the lightcurve and the
corresponding inclination, and our correction to the radial-velocity
amplitude at face value, the formal uncertainty is small,
$\sim\!0.12\,M_\odot$.

As discussed in \Sref{radinc}, however, the lightcurve modelling
relies on a number of assumptions, especially that there is no heat
transport over the face of the star.  From our conservative
constraints on both the inclination and the mass ratio, we find a
lower limit to the mass of $1.66\,M_\odot$.

Thus, from our work we conclude that \psr\ certainly is more massive
than the canonical $1.35\,M_\odot$, and likely substantially more
massive.  Indeed, it may well be more massive even than PSR~J1614-2230
($1.97\pm0.04$; \citealt{demo+10}), and thus allow even more stringent
constraints on the equation of state.  The large mass also suggests a
large amount of mass was transferred in the preceding phase as an
X-ray binary, although this conclusion depends on the initial mass.
However, even if that were as high as the mass found for Vela~X-1
($\sim\!1.9\,M_\odot$, \citealt{barz+01}), our measurements suggest
the pulsar has accreted about half a solar mass.\footnote{Since Vela
  X-1 has a massive companion, its mass must still be close to the one
  it formed with.}

To confirm the high inferred mass will require more secure constraints
on the orbital inclination and, to a lesser extent, the correction
factor for the radial-velocity amplitude.  For this purpose, most
important would be to model the lightcurve in more bands, and to check
explicitly what inclinations are possible for less-constrained,
perhaps even arbitrary temperature distributions.  It may be
especially valuable to model the spectra at the same time, thus
avoiding the indirect calculation of a correction factor.  We are
currently working on rewriting our code for this purpose.

An improved, nearly model-independent constraint on the mass ratio
could be obtained from a near-infrared radial-velocity curve.  Since
in the near-infrared the contribution of the cold side to the light
budget is more important, the radial-velocity amplitude of the center
of light at infrared wavelengths should be much closer to that of the
companion's center of mass, and hence the uncertainty in the
corrections much less important.

Furthermore, one could determine the projected rotational velocity
$v\sin i$ from high-resolution spectra.  This would allow one to check
the predictions from the models, in particular for the mass ratio and
the filling factor, on which $v\sin i$ depends most strongly.

Finally, radio observations could provide a complementary improved
constraint on the inclination, from mapping the eclipse at a larger
range of frequencies than was done by \cite{frucg92}, and making use
of the large increases in sensitivity, especially at high frequency,
that have been made over the last decades.

\acknowledgements The data presented herein were obtained at the
W.M. Keck Observatory, which is operated as a scientific partnership
among the California Institute of Technology, the University of
California and the National Aeronautics and Space Administration. The
Observatory was made possible by the generous financial support of the
W.M. Keck Foundation.  We also used data from the UVES Paranal
Observatory Project UVESPOP (ESO DDT Program ID 266.D-5655).  We made
extensive use of SIMBAD and ADS.

{\it Facilities:} \facility{Keck (LRIS)}

\bibliography{rv}

\end{document}